\pdfoutput=1
\documentclass[12pt,a4paper]{article}
\usepackage{ifthen} % for conditional statements
\newboolean{pdflatex}
\setboolean{pdflatex}{true} % False for eps figures 

\usepackage{comment}
\usepackage{subcaption}

\newboolean{articletitles}
\setboolean{articletitles}{true} % False removes titles in references

\newboolean{uprightparticles}
\setboolean{uprightparticles}{false} %True for upright particle symbols

\def\paperauthors{LHCb collaboration} % Leave as is for PAPER, CONF and FIGURE
\def\paperasciititle{Measurement of the average transverse momentum of forward prompt charged particles in pp and pA collisions at sqrt(s_NN) = 5.02 TeV} % Set ASCII title here !! MAKE sure it's only ASCII characters !! 
\def\papertitle{Measurement of the average transverse momentum of forward prompt charged particles in \pp and \pA collisions at $\sqrt{s_{\mathrm{NN}}}=5.02\tev$} % Latex formatted title
\def\paperkeywords{{High Energy Physics}, {LHCb}} % Comma separated list
\def\papercopyright{\the\year\ CERN for the benefit of the LHCb collaboration} % new since 9/Apr/2018
\def\paperlicence{CC BY 4.0 licence}
\def\paperlicenceurl{https://creativecommons.org/licenses/by/4.0/}

\newif\ifEnableSectionTOCLinks
\EnableSectionTOCLinksfalse % deactivated

\usepackage[top=1in, bottom=1.25in, left=1in, right=1in]{geometry}

\usepackage{microtype}
\usepackage{lineno}  % for line numbering during review
\usepackage{xspace} % To avoid problems with missing or double spaces after
\usepackage{caption} %these three command get the figure and table captions automatically small

\usepackage{graphicx}  % to include figures (can also use other packages)
\usepackage{color}
\usepackage{colortbl}
\graphicspath{{./figs/}} % Make Latex search fig subdir for figures
\usepackage{amsmath} % Adds a large collection of math symbols
\usepackage{amssymb}
\usepackage{amsfonts}
\usepackage{upgreek} % Adds in support for greek letters in roman typeset

\newcommand*\patchAmsMathEnvironmentForLineno[1]{%
\expandafter\let\csname old#1\expandafter\endcsname\csname #1\endcsname
\expandafter\let\csname oldend#1\expandafter\endcsname\csname
end#1\endcsname
 \renewenvironment{#1}%
   {\linenomath\csname old#1\endcsname}%
   {\csname oldend#1\endcsname\endlinenomath}%
}
\newcommand*\patchBothAmsMathEnvironmentsForLineno[1]{%
  \patchAmsMathEnvironmentForLineno{#1}%
  \patchAmsMathEnvironmentForLineno{#1*}%
}
\AtBeginDocument{%
\patchBothAmsMathEnvironmentsForLineno{equation}%
\patchBothAmsMathEnvironmentsForLineno{align}%
\patchBothAmsMathEnvironmentsForLineno{flalign}%
\patchBothAmsMathEnvironmentsForLineno{alignat}%
\patchBothAmsMathEnvironmentsForLineno{gather}%
\patchBothAmsMathEnvironmentsForLineno{multline}%
\patchBothAmsMathEnvironmentsForLineno{eqnarray}%
}

\usepackage[pdftex,
            pdfauthor={\paperauthors},
            pdftitle={\paperasciititle},
            pdfkeywords={\paperkeywords}]{hyperref}
\usepackage{hyperxmp}
\hypersetup{
    pdfcopyright={Copyright (C) \papercopyright},
    pdflicenseurl={\paperlicenceurl}
}
\usepackage[colorinlistoftodos,textsize=scriptsize]{todonotes}

\usepackage[bottom,flushmargin,hang,multiple]{footmisc}

\usepackage[all]{hypcap} % Internal hyperlinks to floats.

\usepackage{xspace} 
\usepackage{upgreek}

\def\lhcb   {\mbox{LHCb}\xspace}

\def\alice  {\mbox{ALICE}\xspace}

\def\MagUp {\mbox{\em Mag\kern -0.05em Up}\xspace}

\ifthenelse{\boolean{uprightparticles}}%
{

 \def\Pmu         {\ensuremath{\upmu}\xspace}

 \def\Ppsi        {\ensuremath{\uppsi}\xspace}

 \def\PDelta      {\ensuremath{\Delta}\xspace}                 
 \def\PXi         {\ensuremath{\Xi}\xspace}                 
 \def\PLambda     {\ensuremath{\Lambda}\xspace}                 
 \def\PSigma      {\ensuremath{\Sigma}\xspace}                 
 \def\POmega      {\ensuremath{\Omega}\xspace}                 
 \def\PUpsilon    {\ensuremath{\Upsilon}\xspace}
 \let\oldPi\Pi
 \def\PPi         {\ensuremath{\oldPi}\xspace}

 \def\PB      {\ensuremath{\mathrm{B}}\xspace}                 
 \def\PD      {\ensuremath{\mathrm{D}}\xspace}                 
 \def\PJ      {\ensuremath{\mathrm{J}}\xspace}                 
 \def\PK      {\ensuremath{\mathrm{K}}\xspace}                 
 \def\Ps      {\ensuremath{\mathrm{s}}\xspace}

 \def\thebaroffset{0.0em}
}
{

 \def\Pmu         {\ensuremath{\mu}\xspace}

 \def\Ppsi        {\ensuremath{\psi}\xspace}                 
                  
 \mathchardef\PDelta="7101
 \mathchardef\PXi="7104
 \mathchardef\PLambda="7103
 \mathchardef\PSigma="7106
 \mathchardef\POmega="710A
 \mathchardef\PUpsilon="7107
 \mathchardef\PPi="7105
 \def\PB      {\ensuremath{B}\xspace}                 
 \def\PD      {\ensuremath{D}\xspace}                 
 \def\PJ      {\ensuremath{J}\xspace}                 
 \def\PK      {\ensuremath{K}\xspace}                 
 \def\Ps      {\ensuremath{s}\xspace}

 \def\thebaroffset{0.18em}
}
\newcommand{\offsetoverline}[2][\thebaroffset]{\kern #1\overline{\kern -#1 #2}}%

\makeatletter
\ifcase \@ptsize \relax% 10pt
  \newcommand{\miniscule}{\@setfontsize\miniscule{4}{5}}% \tiny: 5/6
\or% 11pt
  \newcommand{\miniscule}{\@setfontsize\miniscule{5}{6}}% \tiny: 6/7
\or% 12pt
  \newcommand{\miniscule}{\@setfontsize\miniscule{5}{6}}% \tiny: 6/7
\fi
\makeatother

\DeclareRobustCommand{\optbar}[1]{\shortstack{{\miniscule (\rule[.5ex]{1.25em}{.18mm})}
  \\ [-.7ex] $#1$}}

\def\mup        {{\ensuremath{\Pmu^+}}\xspace}
\def\mun        {{\ensuremath{\Pmu^-}}\xspace} % muon negative (\mum is taken)

\def\squark    {{\ensuremath{\Ps}}\xspace}

\def\kaon    {{\ensuremath{\PK}}\xspace}
\def\KorKbar {\kern \thebaroffset\optbar{\kern -\thebaroffset \PK}{}\xspace}

\def\KS      {{\ensuremath{\kaon^0_{\mathrm{S}}}}\xspace}

\def\D       {{\ensuremath{\PD}}\xspace}

\def\DorDbar {\kern \thebaroffset\optbar{\kern -\thebaroffset \PD}\xspace}

\def\Dp      {{\ensuremath{\D^+}}\xspace}
\def\Dm      {{\ensuremath{\D^-}}\xspace}

\def\DpDm    {\ensuremath{\Dp {\kern -0.16em \Dm}}\xspace}

\def\B       {{\ensuremath{\PB}}\xspace}

\def\BorBbar {\kern \thebaroffset\optbar{\kern -\thebaroffset \PB}\xspace}

\def\Bd      {{\ensuremath{\B^0}}\xspace}

\def\BdorBdbar {\kern \thebaroffset\optbar{\kern -\thebaroffset \Bd}\xspace}

\def\Bs      {{\ensuremath{\B^0_\squark}}\xspace}

\def\BsorBsbar {\kern \thebaroffset\optbar{\kern -\thebaroffset \Bs}\xspace}

\def\jpsi     {{\ensuremath{{\PJ\mskip -3mu/\mskip -2mu\Ppsi}}}\xspace}

\def\Y#1S{\ensuremath{\PUpsilon{(#1S)}}\xspace}

\def\LorLbar     {\kern \thebaroffset\optbar{\kern -\thebaroffset \PLambda}\xspace}
\def\Lambdares   {{\ensuremath{\PLambda}}\xspace}

\newcommand{\decay}[2]{\ensuremath{\mathinner{#1\!\to #2}}\xspace}

\def\to                 {\ensuremath{\rightarrow}\xspace}

\def\AT#1     {\ensuremath{A_{\mathrm{T}}^{#1}}\xspace}           % 2

\def\C#1      {\ensuremath{\mathcal{C}_{#1}}\xspace}                       % 9
\def\Cp#1     {\ensuremath{\mathcal{C}_{#1}^{'}}\xspace}                    % 7
\def\Ceff#1   {\ensuremath{\mathcal{C}_{#1}^{\mathrm{(eff)}}}\xspace}        % 9  
\def\Cpeff#1  {\ensuremath{\mathcal{C}_{#1}^{'\mathrm{(eff)}}}\xspace}       % 7
\def\Ope#1    {\ensuremath{\mathcal{O}_{#1}}\xspace}                       % 2
\def\Opep#1   {\ensuremath{\mathcal{O}_{#1}^{'}}\xspace}                    % 7

\newcommand{\nospaceunit}[1]{\ensuremath{\text{#1}}}       
\newcommand{\aunit}[1]{\ensuremath{\text{\,#1}}}       
\newcommand{\tev}{\aunit{Te\kern -0.1em V}\xspace}
\newcommand{\gev}{\aunit{Ge\kern -0.1em V}\xspace}
\newcommand{\mev}{\aunit{Me\kern -0.1em V}\xspace}
\newcommand{\kev}{\aunit{ke\kern -0.1em V}\xspace}
\newcommand{\ev}{\aunit{e\kern -0.1em V}\xspace}
 
\newcommand{\mevc}{\ensuremath{\aunit{Me\kern -0.1em V\!/}c}\xspace}
\newcommand{\gevc}{\ensuremath{\aunit{Ge\kern -0.1em V\!/}c}\xspace}
\newcommand{\mevcc}{\ensuremath{\aunit{Me\kern -0.1em V\!/}c^2}\xspace}
\newcommand{\gevcc}{\ensuremath{\aunit{Ge\kern -0.1em V\!/}c^2}\xspace}
\def\mm   {\aunit{mm}\xspace}

\def\nb {\aunit{nb}\xspace}
\def\invnb {\ensuremath{\nb^{-1}}\xspace}
\def\pb {\aunit{pb}\xspace}
\def\invpb {\ensuremath{\pb^{-1}}\xspace}

\newcommand{\chisq}{\ensuremath{\chi^2}\xspace}

\def\deriv {\ensuremath{\mathrm{d}}}

\def\gsim{{~\raise.15em\hbox{$>$}\kern-.85em
          \lower.35em\hbox{$\sim$}~}\xspace}
\def\lsim{{~\raise.15em\hbox{$<$}\kern-.85em
          \lower.35em\hbox{$\sim$}~}\xspace}

\def\sqsnn {\ensuremath{\protect\sqrt{s_{\scriptscriptstyle\text{NN}}}}\xspace}
\def\pt         {\ensuremath{p_{\mathrm{T}}}\xspace}

\def\evtgen     {\mbox{\textsc{EvtGen}}\xspace}

\def\geant      {\mbox{\textsc{Geant4}}\xspace}

\def\pythia     {\mbox{\textsc{Pythia}}\xspace}

\def\tell1  {TELL1\xspace}
\def\ukl1   {UKL1\xspace}

\newcommand{\lhcborcid}[1]{\href{https://orcid.org/#1}{\hspace*{0.1em}\raisebox{-0.45ex}{\includegraphics[width=1em]{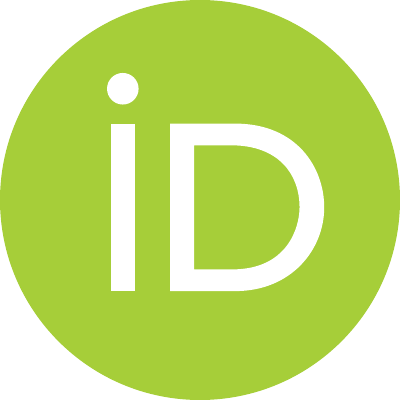}}}}

\input{LHCb/local-symbols-def} % Add in the predefined LHCb symbols

\hypersetup{
  colorlinks   = true, %Colours links instead of ugly boxes
  urlcolor     = blue, %Colour for external hyperlinks
  linkcolor    = blue, %Colour of internal links
  citecolor    = red   %Colour of citations
}

\ifEnableSectionTOCLinks
    \usepackage[explicit]{titlesec} % to change headings
    
    \let\oldcontentsline\contentsline
    \renewcommand

    \titleformat{\section}{\normalfont\Large\bf}{\hyperlink{tocsection.\thesection}{{\thesection} \parbox[t]{\dimexpr\textwidth-1pc}{#1}}}{1pc}{}

    \titleformat{\subsection}{\normalfont\bf}{\hyperlink{tocsubsection.\thesubsection}{{\thesubsection} \parbox[t]{\dimexpr\textwidth-1pc}{#1}}}{1pc}{}

    \titleformat{name=\section,numberless}[display]{}{}{0pt}{\normalfont\Huge\bfseries #1}
\fi

\usepackage{cite} % Allows for ranges in citations
\usepackage{LHCb/mciteplus}
\usepackage{longtable} % only for template; not usually to be used in PAPERs

\begin{document}

%%%%%%%%%%%%%%%%%%%%%%%%%
%%%%% Title     %%%%%%%%%
%%%%%%%%%%%%%%%%%%%%%%%%%
\renewcommand{\thefootnote}{\fnsymbol{footnote}}
\setcounter{footnote}{1}

% %%%%%%% CHOOSE TITLE PAGE--------
%\onecolumn
% ===============================================================================
% Purpose: LHCb-PAPER journal paper title page template
% Author: 
% Created on: 2010-09-25
% ===============================================================================

%%%%%%%%%%%%%%%%%%%%%%%%%
%%%%%  TITLE PAGE  %%%%%%
%%%%%%%%%%%%%%%%%%%%%%%%%
\begin{titlepage}
\pagenumbering{roman}

% Header ---------------------------------------------------
\vspace*{-1.5cm}
\centerline{\large EUROPEAN ORGANIZATION FOR NUCLEAR RESEARCH (CERN)}
\vspace*{1.5cm}
\noindent
\begin{tabular*}{\linewidth}{lc@{\extracolsep{\fill}}r@{\extracolsep{0pt}}}
\ifthenelse{\boolean{pdflatex}}% Logo format choice
{\vspace*{-1.5cm}\mbox{\!\!\!\includegraphics[width=.14\textwidth]{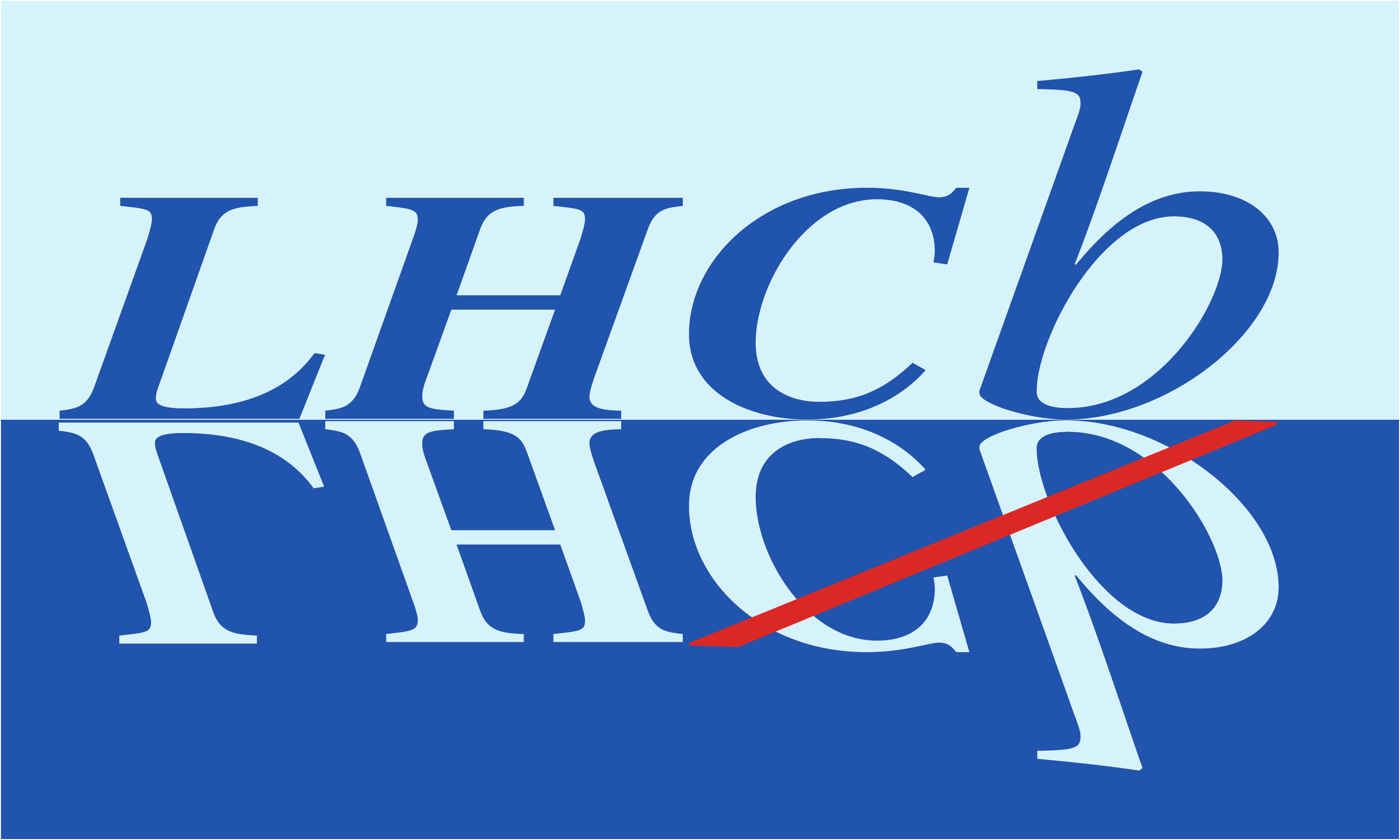}} & &}%
{\vspace*{-1.2cm}\mbox{\!\!\!\includegraphics[width=.12\textwidth]{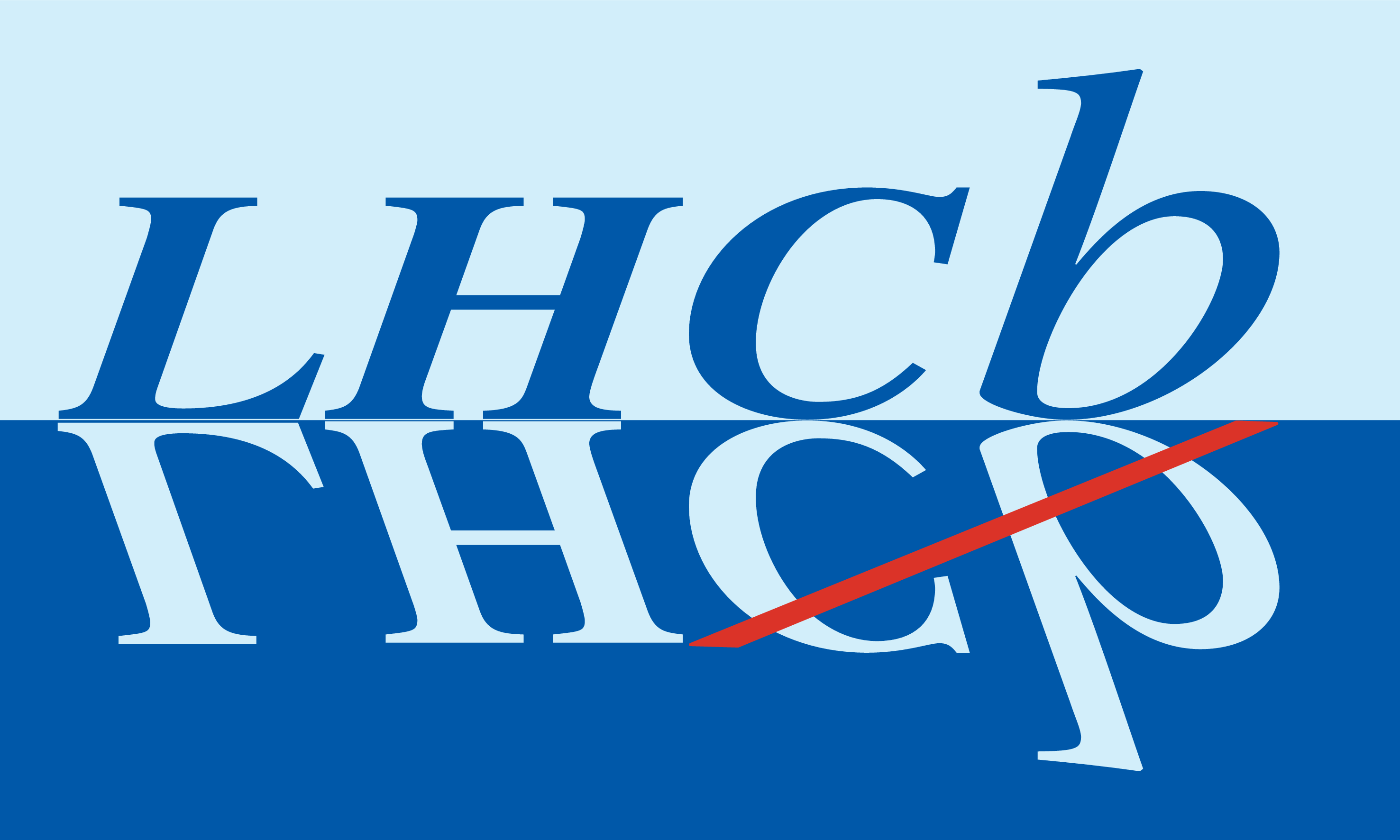}} & &}%
\\
 & & CERN-EP-2026-177 \\  % ID 
 & & LHCb-PAPER-2026-012 \\  % ID 
 & & July 28, 2026 \\ % Date - Can also hardwire e.g.: 23 March 2010
 & & \\
% not in paper \hline
\end{tabular*}

\vspace*{4.0cm}

% Title --------------------------------------------------
{\normalfont\bfseries\boldmath\huge
\begin{center}
% DO NOT EDIT HERE. Instead edit macro in main.tex to keep metadata correct
  \papertitle 
\end{center}
}

\vspace*{2.0cm}

% Authors -------------------------------------------------
\begin{center}
%In the footnote, replace 'paper' by 'Letter' in case of submission to PRL or PLB 
% Edit macro in main.tex to keep metadata correct
\paperauthors\footnote{Authors are listed at the end of this Letter.}
\end{center}

\vspace{\fill}

% Abstract -----------------------------------------------
\begin{abstract}
  \noindent 
 This letter presents the first measurements of the average transverse momentum of prompt charged particles in \pA and \pp collisions as a function of collision multiplicity and pseudorapidity. The data were recorded at nucleon-nucleon centre-of-mass energy $\sqsnn = 5.02\tev$ with the LHCb experiment. The pseudorapidity dependence of the multiplicity distribution is also measured. The average transverse momentum results show a decreasing trend with pseudorapidity, more pronounced in high-multiplicity events, consistent with the collective behaviour of the produced matter. The measurements are reproduced by state-of-the-art (3+1D) hydrodynamic calculations, while saturation models are not compatible with the data.
\end{abstract}

\vspace*{2.0cm}

\begin{center}
  Submitted to Phys.~Rev.~Lett.
\end{center}

\vspace{\fill}

{\footnotesize 
% Edit macro in main.tex to keep metadata correct
\centerline{\copyright~\papercopyright. \href{\paperlicenceurl}{\paperlicence}.}}
\vspace*{2mm}

\end{titlepage}

%%%%%%%%%%%%%%%%%%%%%%%%%%%%%%%%
%%%%%  EOD OF TITLE PAGE  %%%%%%
%%%%%%%%%%%%%%%%%%%%%%%%%%%%%%%%

%  empty page follows the title page ----
\newpage
\setcounter{page}{2}
\mbox{~}
%\newpage
%
%% Author List ----------------------------
%%  You need to get a new author list!
%\input{LHCb_authorlist.tex}
%
%The author list for journal publications is provided by the Membership Committee shortly after 'approval to go to paper' has been given.
%%It will be made available on the page
%%\verb!http://www.physik.uzh.ch/~strauman/forMemCo/LHCb-PAPER-XXXX-XXX/! .
%It will be sent to you by email shortly after a paper number has beens assigned.
%The author list should be included already at first circulation, 
%to allow new members of the collaboration to verify whether they have been included correctly.
%Occasionally a misspelled name is corrected or associated institutions become full members.
%In that case, a new author list will be sent to you.
%In case line numbering doesn't work well after including the authorlist, try moving the \verb!\bigskip! after the last author to a separate line.
%
%
%The authorship for Conference Reports should be ``The LHCb
%  collaboration'', with a footnote giving the name(s) of the contact
%  author(s), but without the full list of collaboration names.

%\twocolumn
% %%%%%%%%%%%%% ---------

\renewcommand{\thefootnote}{\arabic{footnote}}
\setcounter{footnote}{0}

%%%%%%%%%%%%%%%%%%%%%%%%%%%%%%%%
%%%%%  Table of Content   %%%%%%
%%%%%%%%%%%%%%%%%%%%%%%%%%%%%%%%
%%%% Uncomment if desired
%\tableofcontents

\cleardoublepage

%%%%%%%%%%%%%%%%%%%%%%%%%
%%%%% Main text %%%%%%%%%
%%%%%%%%%%%%%%%%%%%%%%%%%

\pagestyle{plain} % restore page numbers for the main text
\setcounter{page}{1}
\pagenumbering{arabic}

%% Uncomment during review phase. 
%% Comment before a final submission.
%\linenumbers

%% This is the main body
%% It is useful to have a single file so comments are not missed in overleaf.
In a broad sense, collectivity refers to phenomena that emerge as the size and complexity of a physical system increase~\cite {Small_systems_review_2024_JeanFiette}. Such behaviour poses important challenges for physics, as its characterisation often requires dedicated experimental techniques and theoretical frameworks. A central question is how collective dynamics in many-body systems arise from the underlying microscopic degrees of freedom, which are typically described as an incoherent superposition of elementary interactions. 

Signatures of collective behaviour, such as azimuthal anisotropies, radial flow, and hadron production yields, have been observed in ultra-relativistic nucleus–nucleus collisions~\cite{Flow_AuAu_STAR, PHENIX_Nature, SHM_Nature}. These observations provide the foundation for interpreting these systems as forming a quark-gluon plasma~(QGP) with near-perfect fluid behaviour. Strikingly, over the past decade, similar signatures of collectivity have also been observed \cite{CMS:2010ifv, LHCb-PAPER-2025-003, ALICE_NATURE_STRANGENESS} in much smaller collision systems, including proton-proton, \pp, and proton-lead, \pA, where collective effects were traditionally expected to be negligible or absent~\cite{Small_systems_review_2024_JeanFiette, Small_Systems_AnnReview}. 

At high energies and forward rapidities in particular, particle production probes the parton distribution functions at small Bjorken-$x$, where the gluon density is very high. This is expected to give rise to a novel state of matter, the colour glass condensate~(CGC), in which the gluon parton distribution functions saturate and nonlinear coherence effects become dominant~\cite{CGC}.
Within the CGC, azimuthal anisotropies and long-range rapidity correlations arise naturally due to initial-state effects from the correlations of gluons already present in the nucleon and the nuclear wavefunctions~\cite{CGC_flow1, CGC_flow2}. However, azimuthal anisotropies that have been observed in heavy-ion collisions were also proposed as a signature for hydrodynamic evolution of a dense, short-lived medium formed in these collisions~\cite{JY_flow_original}.

Due to the absence of a phenomenologically adequate description of collectivity across all system sizes, the study of new observables is crucial. An example of this is the average of the momentum component transverse to the beam direction of charged particles, \meanpt, which has been proposed to distinguish between the two potentially dominant mechanisms: the CGC and the hydrodynamic expansion of the medium~\cite{Bozek_mean_pt, Navarra_mean_pt}. Because of its forward geometry, the~\lhcb experiment is particularly well suited for studies of small-$x$ dynamics and the rapidity dependence of collective behaviour.

This Letter reports the first measurement of \meanpt as a function of the prompt charged particle yield, \Nch, and pseudorapidity, $\eta$, in \pA and \pp collisions at a nucleon-nucleon centre-of-mass energy $\sqsnn = 5.02\tev$ over the \pt range ${0.5 < \pt < 8.0} \gevc$. The charged-particle multiplicity distribution, $P(\Nch)$, is also measured in $\eta$ intervals for the same \pt range. The observables are measured as
\begin{equation}
    \meanptNcheta = \frac{\sum_{\pt} \frac{\deriv^{2} \Nch}{\deriv \pt \deriv\eta} \cdot \pt \cdot \Delta \pt}{\sum_{\pt} \frac{\deriv^{2}\Nch}{\deriv \pt \deriv \eta}\cdot \Delta \pt} \;\text{, where}\; \Nch = \frac{\Ncand \cdot \Pi}{\varepsilon_{\mathrm{reco}}\cdot\varepsilon_{\mathrm{sel}}}.
    \label{eq:<pT>(Nch)}
\end{equation}
The multiplicity distribution is measured as 

\begin{equation}
     P(\Nch)= \frac{1}{N_{\mathrm{evt}}^{\mathrm{T}}} \frac{\deriv  N_{\mathrm{evt}}}{\deriv \Nch}.
    \label{eq:P(Nch)}
\end{equation}
Here, $\Delta \pt$ is the size of the $\pt$ interval, $N_{\rm evt}^{\rm T}$ is the total number of events in the sample, $\deriv N_{\rm evt}/\deriv \Nch$ the number of events in a multiplicity interval, $\Ncand$ the number of prompt charged particle candidates after selection, $\Pi$ the selected prompt charged particles purity, and $\varepsilon_{\mathrm{reco}}$ and $\varepsilon_{\mathrm{sel}}$ are the reconstruction and selection efficiencies. 

The pseudorapidity coverage in the laboratory frame, $\eta_{\mathrm{Lab}}$, is $2.0 < \eta_{\mathrm{Lab}} < 4.8$. For the ion sample, there is a rapidity shift of $+0.465$ $(-0.465)$ for \pA (\Ap).\footnote{When the momentum of a particle is much greater than its mass, its rapidity is approximately equal to
its pseudorapidity ($y \sim \eta$).}

The~\lhcb detector is a single-arm forward spectrometer, described in Refs.~\cite{LHCb-DP-2008-001, LHCb-DP-2014-002}. The subsystems most relevant to this analysis include the silicon-strip vertex locator (VELO) surrounding the interaction region and determining the position of the primary collision vertex~(PV), a tracking system that measures the momenta of charged particles, and two ring-imaging Cherenkov detectors used for charged particle identification.

The analysed \pA (\Ap) data samples were acquired in 2013 and correspond to an integrated luminosity of $42.7 \pm 1.0\invmub$ ($38.7 \pm 1.0\invmub$)~\cite{LHCb-PAPER-2014-047}. 
In \pA collisions, events are selected using a minimum-bias trigger requiring at least one reconstructed track in the VELO. Offline, the PV position is required to be located within three standard deviations of the mean PV position of the full data set. The \pp data set was acquired in 2015 and corresponds to an integrated luminosity of $3.49 \pm 0.07\invnb$. An unbiased trigger is employed for these events, selecting every leading bunch crossing per bunch train. This strategy minimises the contamination from neighbouring bunches to match the \pA conditions. Only events with a single reconstructed PV are analysed for both the \pp and \pA data.

Simulation is used to determine reconstruction efficiencies, to evaluate the impact of selection criteria and to estimate the contributions of backgrounds. The simulation sample for \pA collisions is generated with \epos~\cite{EPOS_LHC}, while for \pp collisions \pythia~\cite{Sjostrand:2007gs} with a specific LHCb configuration is used~\cite{LHCb-PROC-2010-056}. Particle decays are modelled with \evtgen~\cite{Lange:2001uf}, and their interactions with the detector and its response are simulated using the \geant toolkit~\cite{Agostinelli:2002hh}, as described in Ref.~\cite{Allison:2006ve}.

In this Letter, prompt charged particles are defined as charged hadrons and leptons with mean lifetime $\tau > 0.3 \cdot 10^{-10}\;\mathrm{s}$, produced directly from the collisions or originating from decays of shorter-lifetime particles~\cite{ALICE_charged_part_def}. Prompt charged-particle candidates are defined as tracks with hits in the VELO and in the tracking stations downstream of the~\lhcb dipole magnet, which implies a momentum requirement of ${p > 2\; \gevc }$.

Background contributions from fake tracks, secondary particles and clone tracks are accounted for in this analysis. Fake tracks, which do not correspond to genuine charged particles, are artefacts of the reconstruction algorithms and become particularly significant at high detector occupancies. Secondary particles originate from interactions with the detector material or from decays of prompt particles. Clone tracks correspond to two tracks produced by the same particle.

Selection criteria are applied to suppress these backgrounds. Fake tracks are reduced by imposing a tight requirement on the output of a neural-network-based classifier, the ghost probability~\cite{DeCian:2255039}. 
Further suppression is achieved by retaining only the candidate with the best track-fit quality when multiple tracks in the same event share a segment reconstructed in the VELO. 
Secondary particles are suppressed by requiring an impact parameter, defined as the closest distance between the PV and the track~\cite{LHCb-TDR-013, LHCb-DP-2019-001}, smaller than $0.51\mm$ ($0.57\mm$) for \pA (\pp) collisions. 
This requirement is particularly effective in rejecting hadrons from \KS and \Lambdares decays, as well as particles produced in interactions with the detector material. The residual contributions of fake and secondary tracks after the selection are estimated using simulation and subtracted from the signal yield.

The prompt charged-particle yield distribution in \pA, \Ap, and \pp collisions is determined event-by-event from the number of selected candidates, corrected for reconstruction and selection efficiencies and signal purity. As the measurement is sensitive to the event multiplicity, all efficiencies and backgrounds are computed in intervals of $(\eta, \pt, \velotr)$ and applied track by track. Here $\velotr$ stands for the number of tracks in the VELO, which serves as an event-multiplicity proxy. The reconstruction efficiency accounts for the particles that are not reconstructed in the detector, and is computed from simulation and corrected using a tag-and-probe method in two-dimensional intervals of $\eta$ and $p$. The procedure uses \decay{\jpsi}{\mup\mun} decays with muon momenta in the range $5 < p < 200\gevc$\cite{LHCb-DP-2013-002}. Outside this range, tracks are corrected with simulation, and a systematic uncertainty is assigned. The selection efficiency quantifies the fraction of prompt charged particles removed by the applied selection criteria and is estimated from the simulation.

As the reconstruction efficiency depends on detector occupancy, the simulated samples are weighted to match the occupancy distributions observed in data across the \lhcb tracking detectors. 
This efficiency also depends on the particle species. Particle composition extracted from the simulation is validated using measurements from the 
\alice~\cite{ALICE_fractions, Strangeness_enha_Nature, MultiStrangeBaryon} and \lhcb~\cite{LHCb-PAPER-2011-037} experiments. 
For \pp collisions, the \pythia \lhcb tune does not accurately reproduce the relative abundances of kaons and prompt hyperons at high \pt, which would lead to an overestimation of the prompt charged particle reconstruction efficiency. Consequently, a dedicated \epos sample is used to parametrise the particle composition of \pp collisions. The relative particle abundances from \epos agree with data within 30\% for both \pp and \pA/\Ap collisions~\cite{LHCb-PROC-2010-056, ALICE_identified_charged}.

The purity in Eq.~\ref{eq:P(Nch)} is computed as $\Pi = 1  - (f_{\rm fakes} + f_{\rm sec} + f_{\rm clone}^{\rm sim})$. The clone track fraction, $f_{\rm clone}^{\rm sim}$, is computed in simulation, and the fake, $f_{\rm fakes}$, and secondary fraction, $f_{\rm sec}$, are computed in simulation and multiplied by data/simulation correction factors from the dedicated calibration samples. The fake-track background sample is constructed using two independent fake-track calibration samples. One consists of tracks with a high ghost probability, the other is made up of tracks that share a reconstructed VELO segment with a track of better fit quality. Contributions from secondary particles are evaluated separately for hadrons and electrons. A hadron-calibration sample, consisting of decay products of weakly decaying strange hadrons such as \Lambdares baryons, \KS mesons, or produced in interactions with detector material, is selected using tracks with large impact parameters. The electron contribution from photon conversions, which is significant at low \pt, is estimated initially from simulation with further corrections determined from an electron-dedicated calibration sample selected using the particle identification detectors.

As bin migration in the multiplicity distribution has a strong effect, a  Bayesian unfolding procedure is used~\cite{Adye:2011gm, BayesianUnfolding}. A closure test is performed using simulation to verify that the original distribution is recovered within statistical uncertainties. A closure test is also performed for the \meanpt observable, using simulated events that are processed through the full analysis chain. The discrepancy between the generator-level simulation and the measured data is assigned as a systematic uncertainty on \meanpt due to event multiplicity classification and varies from 1.0 to 4.5\%. Other potential sources of systematic uncertainty are also investigated. For candidates with momenta in the range $5 < p < 200\gevc$, the track-finding efficiency uncertainty arises from the limited size of the \decay{\jpsi}{\mup\mun} calibration samples and from differences between hadron and muon interactions with detector material. For candidates outside this range, the largest uncertainty found in the $5 < p < 200\gevc$ range is assigned, corresponding to 5\%. 
The uncertainty associated with the particle composition in simulation is evaluated by varying the relative particle abundances by 30\%. The purity systematic uncertainty accounts for two sources: sample sizes and the purities of the dedicated calibration samples. A closure test is carried out using simulation to assess potential bias arising from the multiplicity classification. The difference between the simulation treated as pseudodata and the generator-level simulation is assigned as an uncertainty of the event classification procedure. The closure test uncertainty dominates for high-multiplicity events, while the reconstruction efficiency uncertainty dominates for low-multiplicity events. Overall, the total uncertainty on \meanpt is less than 5\% in all multiplicity bins and dominated by systematic uncertainties. Because of high systematic uncertainties, the first pseudorapidity bin is excluded from the \meanpt measurement.

\begin{figure}[bt!]
     \centering
     \begin{subfigure}[t]{0.48\textwidth}
         \centering
         \includegraphics[width=\linewidth]{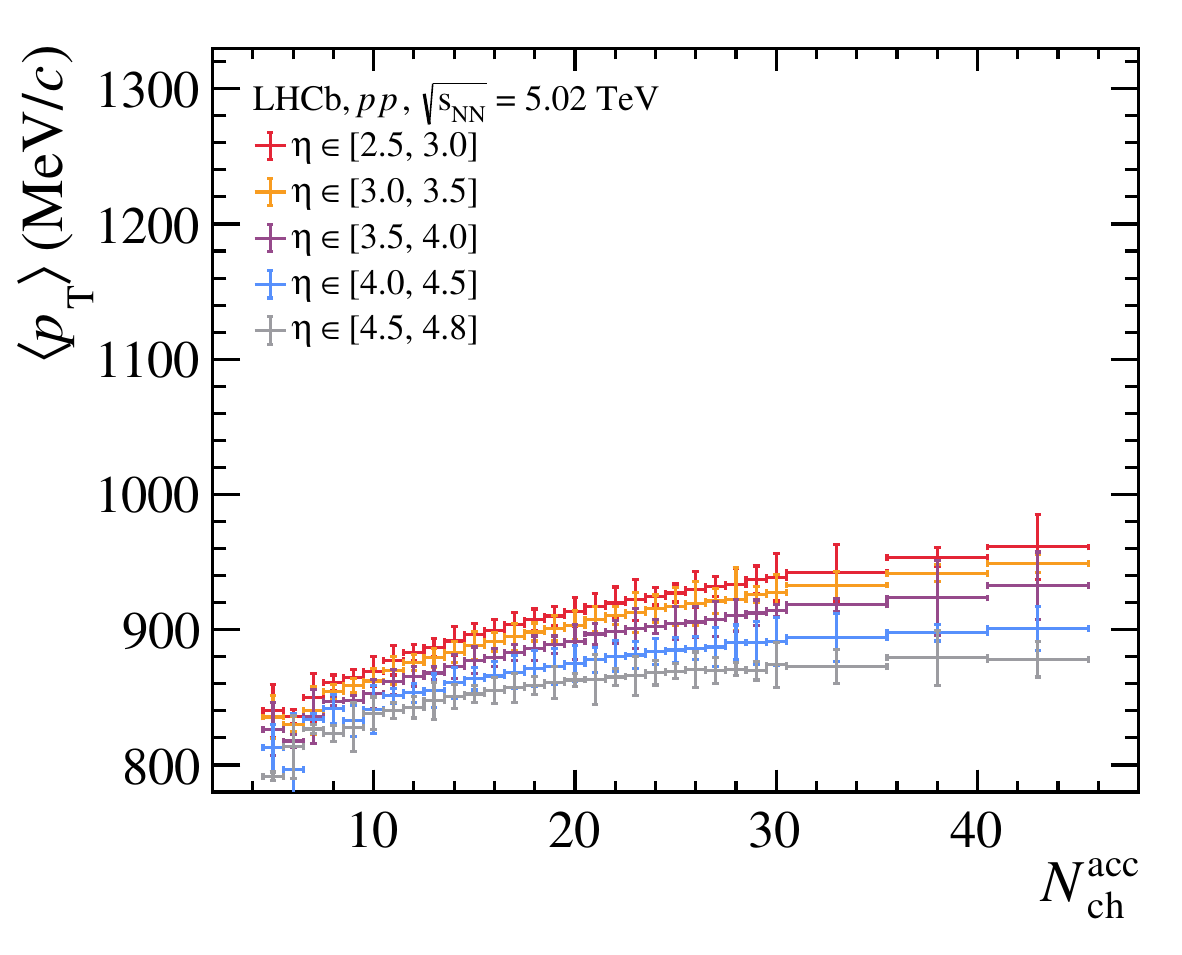}
     \end{subfigure}
     \hfill
     \begin{subfigure}[t]{0.48\textwidth}
         \centering
         \includegraphics[width=\linewidth]{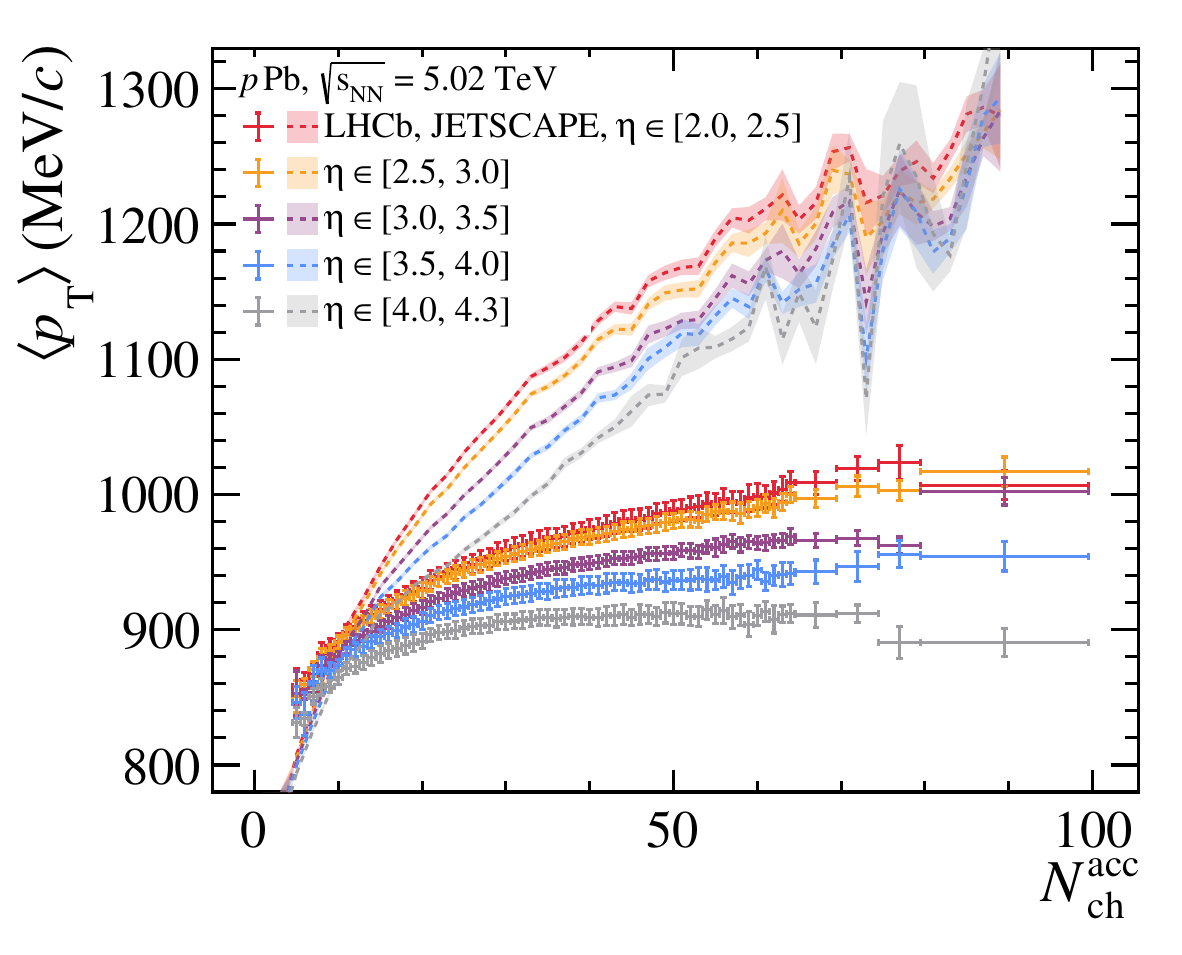}
     \end{subfigure}
     \begin{subfigure}[t]{0.48\textwidth}
         \centering
         \includegraphics[width=\linewidth]{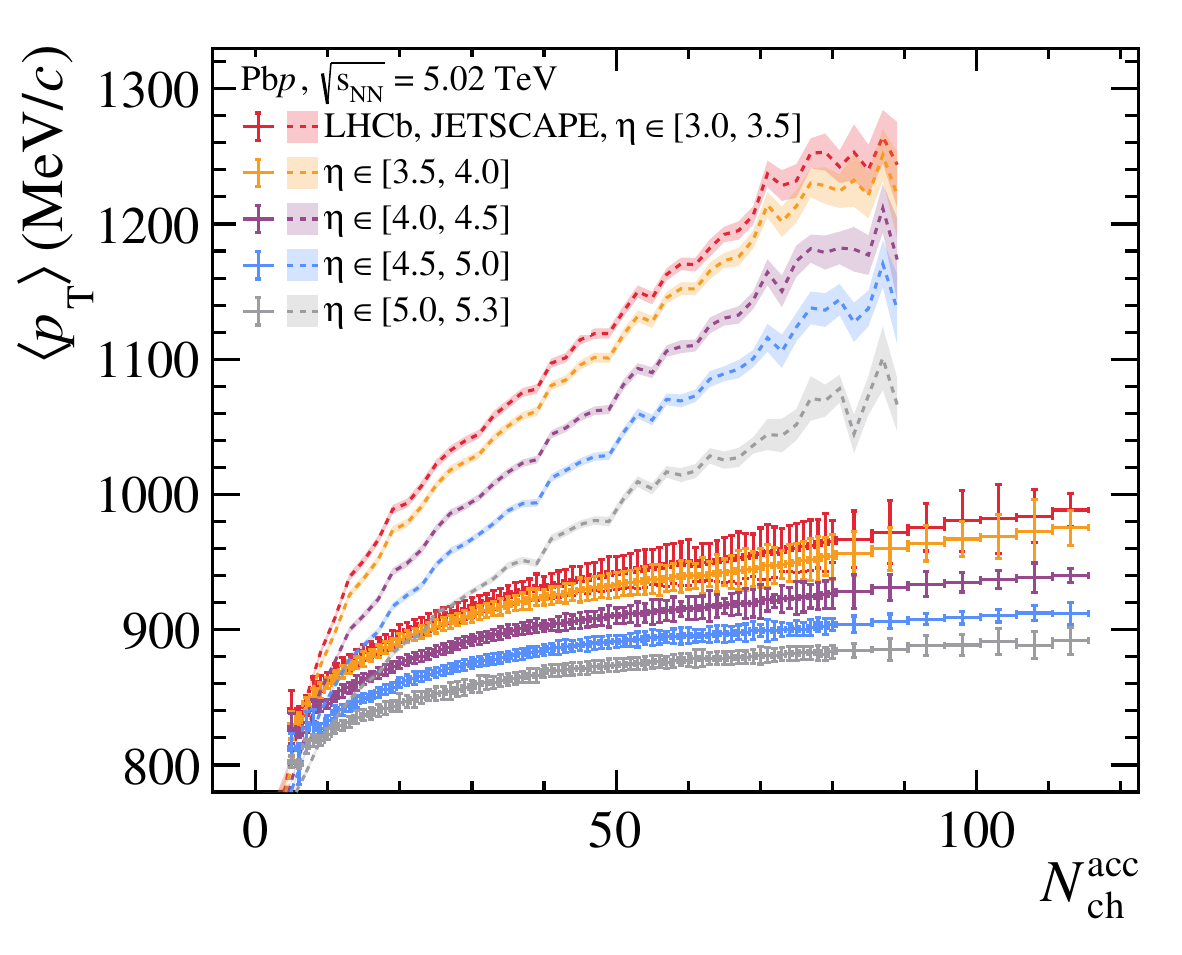} 
     \end{subfigure}
\caption{Distribution of \meanpt as a function of multiplicity and pseudorapidity in the centre-of-mass for \pp (top left), \pA (top right) and \Ap (bottom) collisions. The dashed lines correspond to JETSCAPE calculations using relativistic hydrodynamics~\cite{JETSCAPE}.}
\label{fig:mean_pt_eta_Nch}
\end{figure}

Figure~\ref{fig:mean_pt_eta_Nch} shows that the observed \meanpt increases with the event multiplicity for all three collision systems, which is consistent with observations in larger collision systems by the ALICE collaboration~\cite{ALICE_mean_pt_2013, ALICE_XeXe_mean_pt}. A similar trend is found for all pseudorapidity bins. More forward $\eta$ regions have, in general, smaller \meanpt values than the less forward ones.  This pseudorapidity splitting appears at high multiplicity, while at low multiplicity, all $\eta$ intervals tend to converge to similar \meanpt values. This is consistent with the emergence of collective behaviour that is noted at high-multiplicity events. 

\begin{figure}[bt!]
     \centering
     \begin{subfigure}[t]{0.48\textwidth}
         \centering
         \includegraphics[width=\linewidth]{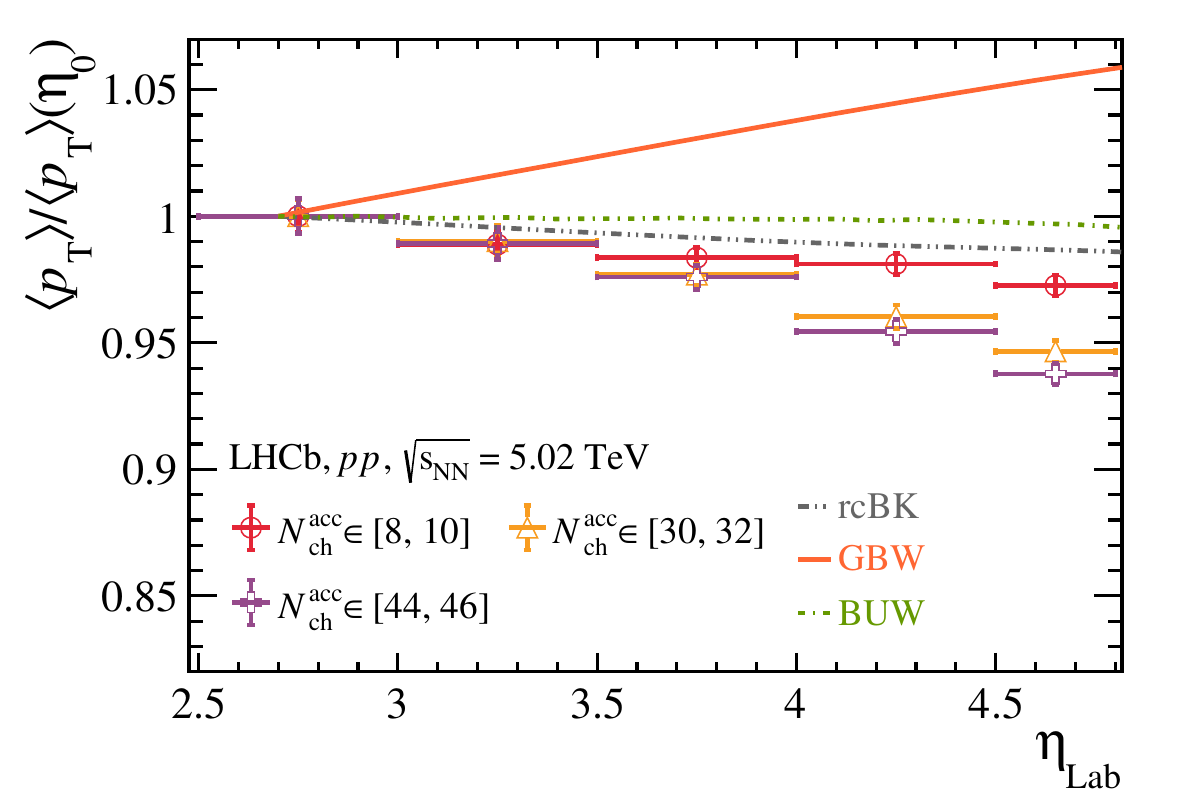}
     \end{subfigure}
     \hfill
     \begin{subfigure}[t]{0.48\textwidth}
         \centering
         \includegraphics[width=\linewidth]{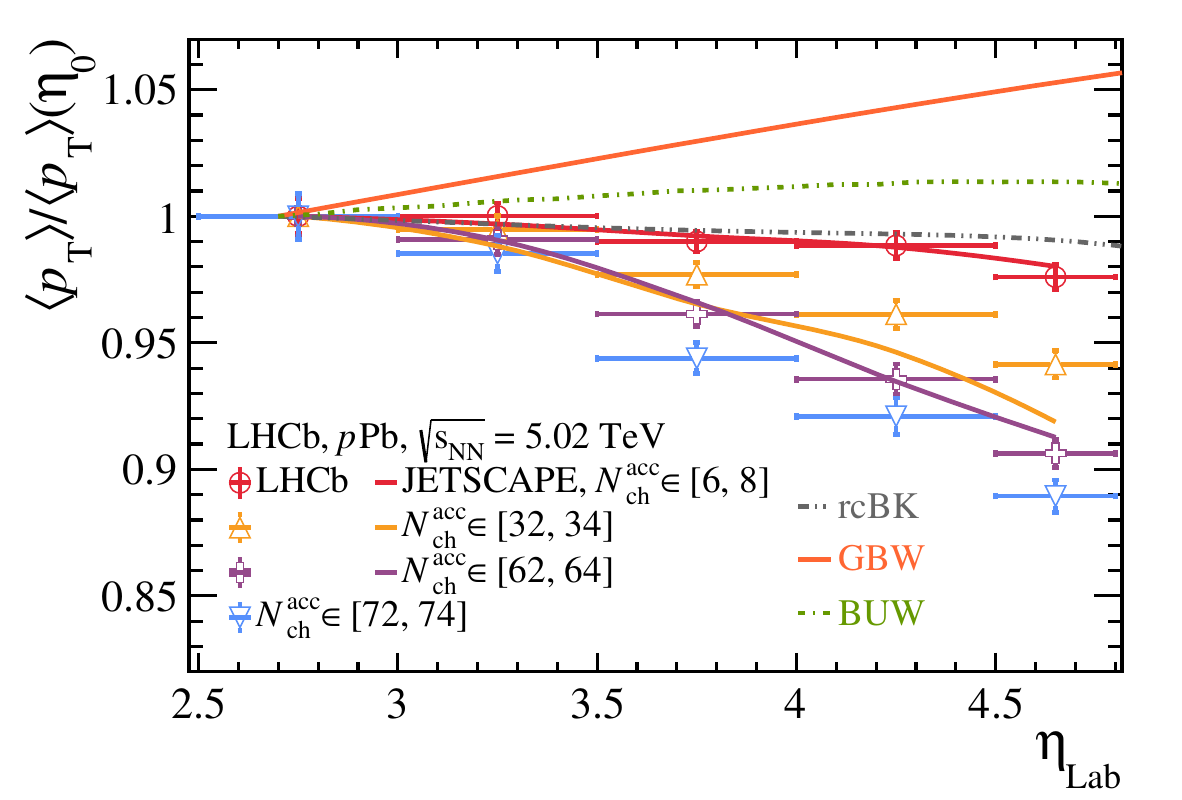}
     \end{subfigure}
     \begin{subfigure}[t]{0.48\textwidth}
         \centering
         \includegraphics[width=\linewidth]{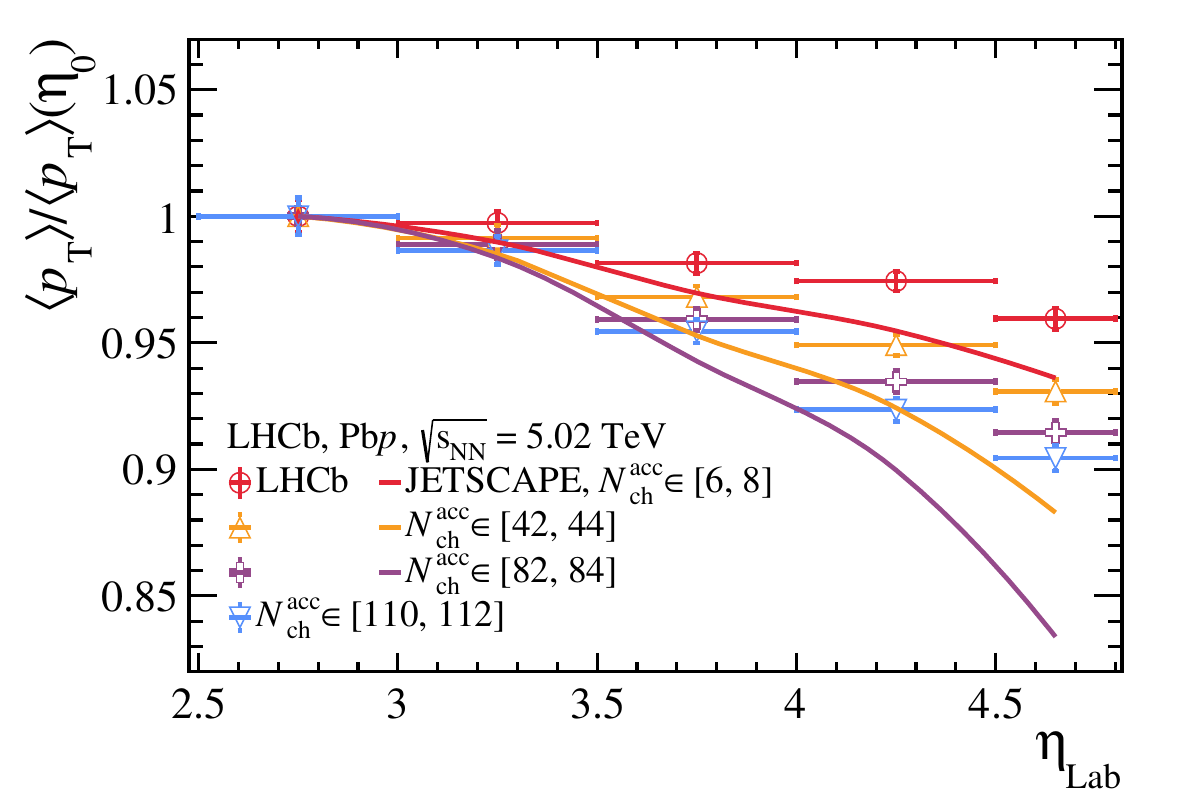}
     \end{subfigure}
\caption{Results of \meanpt normalised to the bin $\eta_{0}\in [2.5, 3.0]$ as a function of $\eta_{\textrm{Lab}}$ for (top left) \pp , (top right) \pA, and  (bottom) \Ap data. The data are compared to various CGC predictions (rcBK, GBW and BUW) obtained from Ref.~\cite{Navarra_mean_pt}, which are not differential in multiplicity. The solid lines are computed by the JETSCAPE collaboration based on Ref.~\cite{JETSCAPE}.}
\label{fig:mean_pt_models}
\end{figure}

In Fig.~\ref{fig:mean_pt_models}, the measured value of \meanpt normalised by the pseudorapidity bin in the laboratory frame $\eta_{0}\in [2.5,\; 3.0]$ is shown. The observed \meanpt decrease in the most forward region is about 5--10\% and can be qualitatively described by a relativistic hydrodynamic calculation from the JETSCAPE collaboration~\cite{JETSCAPE}. 
The results from \pA and \Ap data show similar behaviour, which is consistent with a negligible effect of the Bjorken-$x$ region change as observed in previous flow measurements~\cite{LHCb-PAPER-2025-003}. 
As can be seen in Fig.~\ref{fig:mean_pt_models}, in general, the CGC models do not reproduce the pseudorapidity trend in data. At low multiplicity, the CGC model with running coupling Balitsky-Kovchegov evolution equations, rcBK~\cite{rcBK}, reproduces the data qualitatively in \pp collisions. However, the Golec-Biernat-Wusthoff (GBW) approach~\cite{GBW_model} predicts that \meanpt increases with $\eta$, which is inconsistent with the data. The hydrodynamic calculation from JETSCAPE reproduces the \pA data while it overestimates the decrease in \meanpt in the most backward region of \Ap. These data provide a unique view of collective effects in small collision systems using an observable not measured previously.

\begin{figure}[bt!]
     \centering
     \begin{subfigure}[t]{0.48\textwidth}
         \centering
         \includegraphics[width=\linewidth]{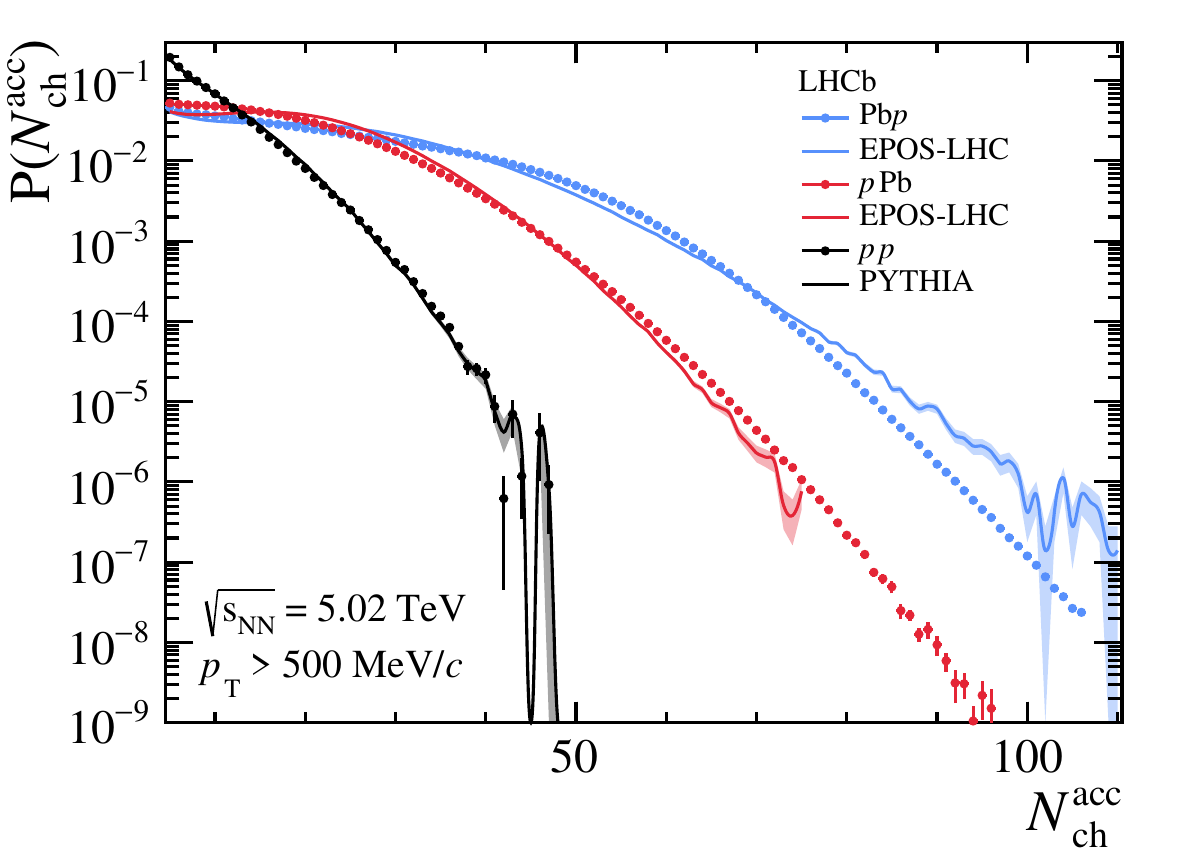}
     \end{subfigure}
\caption{Multiplicity distribution results integrated over pseudorapidity. The comparison with the simulation models \epos~\cite{EPOS_LHC} for \pA and \pythia~\cite{Sjostrand:2007gs} for \pp is also shown. }
\label{fig:PNch_models}
\end{figure}

Finally, the multiplicity distributions are shown in Fig.~\ref{fig:PNch_models}. While comparing the three different systems, it can be observed that \Ap produces larger multiplicities than \pp and \pA collisions. The models are found to describe the data qualitatively. The region that is least well reproduced is the higher-multiplicity region in \Ap collisions, where \epos overestimates the data.

In summary, this Letter presents the first measurement of the average transverse momentum of forward prompt charged particles for $\pp$ and $\pA$/\Ap collisions at $\mbox{\sqsnn = 5.02\tev}$. This measurement is done for the first time as a function of event multiplicity and pseudorapidity. A decreasing dependence of the prompt charged particle average transverse momentum with increasing pseudorapidity is observed for the first time in small collision systems. The data are qualitatively described by JETSCAPE hydrodynamic simulations, while CGC-based models fail to reproduce the data, especially at high multiplicities. The pseudorapidity trend is stronger in high-multiplicity events, consistent with the expectations of collective behaviour.

% Comment this in for paper drafts; do not include this in analysis note, conference and figure reports
\section*{Acknowledgements}
%
% These Acknowledgements valid from 16/06/2026
%
\noindent We express our gratitude to our colleagues in the CERN
accelerator departments for the excellent performance of the LHC. We
thank the technical and administrative staff at the LHCb
institutes.
We acknowledge support from CERN and from the national agencies:
ARC (Australia);
CAPES, CNPq, FAPERJ and FINEP (Brazil); 
MOST and NSFC (China); 
CNRS/IN2P3 and CEA (France);  % added CEA 26/02/2026
BMFTR, DFG and MPG (Germany);
NKFIH (Hungary);              % added 16/06/2026
INFN (Italy); 
NWO (Netherlands); 
MNiSW and NCN (Poland); 
MEC/IFA (Romania); 
%MSHE (Russia); 
MICIU and AEI (Spain);
SNSF and SER (Switzerland); 
NASU (Ukraine); 
STFC (United Kingdom); 
DOE NP and NSF (USA).
%%%%%%%%%%%%%%%%%%%%%%%%%%%%%%%%%%%%%%%%%%%%%
We acknowledge the computing resources that are provided by ARDC (Australia), 
CBPF (Brazil),
CERN, 
IHEP and LZU (China),
IN2P3 (France), 
KIT and DESY (Germany), 
INFN (Italy), 
SURF (Netherlands),
Polish WLCG (Poland),
IFIN-HH (Romania), % http://dx.doi.org/10.13039/100019931,"Institutul National de Cercetare-Dezvoltare pentru Fizica si Inginerie Nucleara 'Horia Hulubei'"
%RRCKI and Yandex LLC (Russia), 
PIC (Spain), CSCS (Switzerland), 
GridPP (United Kingdom),
and NSF (USA).  % added Feb2026
%%%%%%%%%%%%%%%%%%%%%%%%%%%%%%%%%%%%%%%%%% 
We are indebted to the communities behind the multiple open-source
software packages on which we depend.
%%%%%%%%%%%%%%%%%%%%%%%%%%%%%%%%%%%%%%%%%%
Individual groups or members have received support from
% ARC and ARDC (Australia); % moved to national 16/01/2025
RTP (Australia), % added 06/03/2026
FWO Odysseus grant G0ASD25N (Belgium), % added 20/4/2026
Key Research Program of Frontier Sciences of CAS, CAS PIFI, CAS CCEPP (China); 
%Fundamental Research Funds for the Central Universities,  and Sci.\
%\& Tech.\ Program of Guangzhou (China); Removed 24/11/25
Minciencias (Colombia);
EPLANET, Marie Sk\l{}odowska-Curie Actions, ERC and NextGenerationEU (European Union);
A*MIDEX, ANR, IPhU and Labex P2IO, and R\'{e}gion Auvergne-Rh\^{o}ne-Alpes (France);
%RFBR, RSF and Yandex LLC (Russia);
Alexander-von-Humboldt Foundation (Germany);
ICSC (Italy); 
%GVA, XuntaGal, GENCAT, Inditex, InTalent and Prog.~Atracci\'on Talento, CM (Spain);
Severo Ochoa and Mar\'ia de Maeztu Units of Excellence, GVA, XuntaGal, GENCAT, InTalent-Inditex and Prog.~Atracci\'on Talento CM (Spain);
%XuntaGal --> Xunta de Galicia 
% SRC (Sweden);  % removed 27/02/2026 - end of grant
the Leverhulme Trust, the Royal Society and UKRI (United Kingdom).

\addcontentsline{toc}{section}{References}
%\setboolean{inbibliography}{true}
\bibliographystyle{LHCb/LHCb}
\bibliography{main,LHCb/standard,LHCb/LHCb-PAPER,LHCb/LHCb-CONF,LHCb/LHCb-DP,LHCb/LHCb-TDR,LHCb/LHCb-PUB}

@article{LHCb-DP-2019-001,
      author         = "Roel Aaij and others",
      title         = "{Design and performance of the LHCb trigger and full real-time  reconstruction in Run 2 of the LHC}",
      eprint         = "1812.10790",
      archivePrefix  = "arXiv",
      primaryClass   = "hep-ex",
      number         = "LHCb-DP-2019-001",
      year           = "2019",
      journal        = "JINST",
      volume         = "14",
      pages          = "P04013",
      doi            = "10.1088/1748-0221/14/04/P04013",
      }

@article{LHCb-DP-2014-002,
      author         = "Aaij, R. and others",
      title          = "{LHCb detector performance}",
      collaboration  = "LHCb collaboration",
      journal        = "Int. J. Mod. Phys.",
      volume         = "A30",
      pages          = "1530022",
      doi            = "10.1142/S0217751X15300227",
      year           = "2015",
      eprint         = "1412.6352",
      archivePrefix  = "arXiv",
      primaryClass   = "hep-ex",
      report         = "LHCB-DP-2014-002, CERN-PH-EP-2014-290",
}

@article{LHCb-DP-2013-002,
      author         = "Aaij, R. and others",
      title          = "{Measurement of the track reconstruction efficiency at LHCb}",
      collaboration  = "LHCb collaboration",
      journal        = "JINST",
      volume         = "10",
      pages          = "P02007",
      doi            = "10.1088/1748-0221/10/02/P02007",
      year           = "2015",
      eprint         = "1408.1251",
      archivePrefix  = "arXiv",
      primaryClass   = "hep-ex",
      report         = "CERN-LHCB-DP-2013-002",
}

@article{LHCb-DP-2008-001,
      author         = "Alves~Jr., A. A. and others",
      title          = "{The \lhcb detector at the LHC}",
      collaboration  = "LHCb collaboration",
      journal        = "JINST",
      volume         = "3",
      pages          = "S08005",
      doi            = "10.1088/1748-0221/3/08/S08005",
      year           = "2008",
      number         = "LHCb-DP-2008-001",
}

@article{LHCb-PAPER-2025-003,
      author         = "Aaij, R. and others",
      title          = "{Rapidity and multiplicity dependence of charged-particle flow in $p$Pb collisions at $\sqsnn = 8.16$ \tev}",
      collaboration  = "LHCb collaboration",
      report         = "{LHCb-PAPER-2025-003 CERN-EP-2025-090}",
      eprint         = "2505.09273",
      archivePrefix  = "arXiv",
      primaryClass   = "hep-ex",
      journal        = "{JHEP}",
      volume         = "10",
      pages          = "124",
      year           = "2025",
      doi            = "10.1007/JHEP10(2025)124"
}

@article{LHCb-PAPER-2014-047,
      author         = "Aaij, R. and others",
      title          = "{Precision luminosity measurements at LHCb}",
      collaboration  = "LHCb collaboration",
      journal        = "JINST",
      volume         = "9",
      pages          = "P12005",
      doi            = "10.1088/1748-0221/9/12/P12005",
      year           = "2014",
      eprint         = "1410.0149",
      archivePrefix  = "arXiv",
      primaryClass   = "hep-ex",
      report         = "LHCb-PAPER-2014-047 CERN-PH-EP-2014-221",
}

@article{LHCb-PAPER-2011-037,
      author         = "Aaij, R. and others",
      title          = "{Measurement of prompt hadron production ratios in 
                        \proton\proton collisions at $\sqs=0.9$ and 7~\tev}",
      collaboration  = "LHCb collaboration",
      journal        = "Eur. Phys. J.",
      volume         = "C72",
      pages          = "2168",
      doi            = "10.1140/epjc/s10052-012-2168-x",
      year           = "2012",
      eprint         = "1206.5160",
      archivePrefix  = "arXiv",
      primaryClass   = "hep-ex",
      report         = "LHCb-PAPER-2011-037 CERN-PH-EP-2012-171",
}

@lhcbreport{LHCb-TDR-013,
      collaboration = "LHCb collaboration",
      title         = "{LHCb VELO Upgrade Technical Design Report}",
      institution   = "CERN",
      address       = "Geneva",
      number        = "{CERN-LHCC-2013-021}",
      year          = "2013",
}

@article{LHCb-PROC-2010-056,
      author         = "Belyaev, I. and others",
      title          = "{Handling of the generation of primary events
                         in Gauss, the LHCb simulation framework}",
      journal="J. Phys. Conf. Ser.",
      volume={331},
      pages={032047},
      doi={10.1088/1742-6596/331/3/032047},
      year={2011},
}

@article{Sjostrand:2007gs,
      author         = {Sj\"{o}strand, Torbj\"{o}rn and Mrenna, Stephen and
                        Skands, Peter"},
      title          = "{A brief introduction to PYTHIA 8.1}",
      journal        = "Comput. Phys. Commun.",
      volume         = "178",
      pages          = "852-867",
      doi            = "10.1016/j.cpc.2008.01.036",
      year           = "2008",
      eprint         = "0710.3820",
      archivePrefix  = "arXiv",
      primaryClass   = "hep-ph",
      reportNumber   = "CERN-LCGAPP-2007-04, LU-TP-07-28,
                        FERMILAB-PUB-07-512-CD-T",
}

@Article{Agostinelli:2002hh,
     author    = "Agostinelli, S. and others",
 collaboration = "Geant4 collaboration",
     title     = "{Geant4: A simulation toolkit}",
     journal   = "Nucl. Instrum. Meth.",
     volume    = "A506",
     year      = "2003",
     pages     = "250",
     doi       = "10.1016/S0168-9002(03)01368-8",
}

@article{Allison:2006ve,
      author         = "Allison, John and Amako, K. and Apostolakis, J. and
                        Araujo, H. and Dubois, P.A. and others",
 collaboration = "Geant4 collaboration",
      title          = "{Geant4 developments and applications}",
      journal        = "IEEE Trans.Nucl.Sci.",
      volume         = "53",
      pages          = "270",
      doi            = "10.1109/TNS.2006.869826",
      year           = "2006",
      reportNumber   = "SLAC-PUB-11870",
}

@Article{Lange:2001uf,
     author    = "Lange, D. J.",
     title     = "{The EvtGen particle decay simulation package}",
     journal   = "Nucl. Instrum. Meth.",
     volume    = "A462",
     year      = "2001",
     pages     = "152-155",
     doi       = "10.1016/S0168-9002(01)00089-4",
}

@inproceedings{Adye:2011gm,
    author = "Adye, Tim",
    title = "{Unfolding algorithms and tests using RooUnfold}",
    booktitle = "{PHYSTAT 2011}",
    eprint = "1105.1160",
    archivePrefix = "arXiv",
    primaryClass = "physics.data-an",
    doi = "10.5170/CERN-2011-006.313",
    publisher = "CERN",
    address = "Geneva",
    pages = "313--318",
    year = "2011"
}

@lhcbreport{DeCian:2255039,
      author        = "De Cian, Michel and Farry, Stephen and Seyfert, Paul and
                       Stahl, Sascha",
      title         = "{Fast neural-net based fake track rejection in the LHCb
                       reconstruction}",
      institution   = "CERN",
      address       = "Geneva",
      number        = "LHCb-PUB-2017-011",
      year          = "2017",
      reportNumber  = "LHCb-PUB-2017-011",
      url           = "http://cds.cern.ch/record/2255039",
}

@article{PHENIX_Nature,
author="Aidala, C. and others",
title={Creation of quark-gluon plasma droplets with three distinct geometries},
collaboration="PHENIX collaboration",
journal={Nature Phys.},
eprint = "1805.02973",
archivePrefix = "arXiv",
year={2019},
month={Mar},
day={01},
volume={15},
number={3},
pages={214-220},
issn={1745-2481},
doi={10.1038/s41567-018-0360-0},
url={https://doi.org/10.1038/s41567-018-0360-0}
}

@article{Small_Systems_AnnReview,
   title={Small System Collectivity in Relativistic Hadronic and Nuclear Collisions},
   eprint = "1801.03477",
   archivePrefix = "arXiv",
   volume={68},
   ISSN={1545-4134},
   url={http://dx.doi.org/10.1146/annurev-nucl-101916-123209},
   DOI={10.1146/annurev-nucl-101916-123209},
   number={1},
   journal={Ann. Rev. Nucl. Part. Sci.},
   publisher={Annual Reviews},
   author={Nagle, James L. and Zajc, William A.},
   year={2018},
   month=oct, pages={211–235} }

@article{Small_systems_review_2024_JeanFiette,
      title={A Decade of Collectivity in Small Systems}, 
      author={Jan Fiete Grosse-Oetringhaus and Urs Achim Wiedemann},
      year={2024},
      eprint={2407.07484},
      archivePrefix={arXiv},
      primaryClass={hep-ex},
      url={https://arxiv.org/abs/2407.07484}, 
}

@article{CGC,
   title={The {C}olor {G}lass {C}ondensate},
   volume={60},
   eprint = "1002.0333",
   archivePrefix = "arXiv",
   ISSN={1545-4134},
   url={http://dx.doi.org/10.1146/annurev.nucl.010909.083629},
   DOI={10.1146/annurev.nucl.010909.083629},
   number={1},
   journal={Ann. Rev. Nucl. Part. Sci.},
   publisher={Annual Reviews},
   author={Gelis, Francois and Iancu, Edmond and Jalilian-Marian, Jamal and Venugopalan, Raju},
   year={2010},
   month=nov, pages={463–489} }

@article{Bozek_mean_pt,
   title={The rapidity dependence of the average transverse momentum in p+{P}b collisions at the {LHC}: The {C}olor {G}lass {C}ondensate versus hydrodynamics},
   archivePrefix = "arXiv",
   eprint = "1309.7358",
   volume={B728},
   ISSN={0370-2693},
   url={http://dx.doi.org/10.1016/j.physletb.2013.12.034},
   DOI={10.1016/j.physletb.2013.12.034},
   journal={Phys. Lett.},
   publisher={Elsevier BV},
   author={Bożek, Piotr and Bzdak, Adam and Skokov, Vladimir},
   year={2014},
   month=jan, pages={662–665} }

@article{Navarra_mean_pt,
   title={Rapidity dependence of the average transverse momentum in hadronic collisions},
   volume={C94},
   archivePrefix = "arXiv",
   eprint = "1510.04737",
   ISSN={2469-9993},
   url={http://dx.doi.org/10.1103/PhysRevC.94.024917},
   DOI={10.1103/physrevc.94.024917},
   number={2},
   journal={Phys. Rev.},
   publisher={American Physical Society (APS)},
   author={Durães, F. O. and Giannini, A. V. and Gonçalves, V. P. and Navarra, F. S.},
   year={2016},
   month=aug}

@article{EPOS_LHC,
  title = {{EPOS LHC}: Test of collective hadronization with data measured at the {CERN} {L}arge {H}adron {C}ollider},
  author = {Pierog, T. and Karpenko, Iu. and Katzy, J. M. and Yatsenko, E. and Werner, K.},
  journal = {Phys. Rev.},
  archivePrefix = "arXiv",
  eprint = " d1306.0121",
  volume = {C92},
  issue = {3},
  pages = {034906},
  numpages = {15},
  year = {2015},
  month = {Sep},
  publisher = {American Physical Society},
  doi = {10.1103/PhysRevC.92.034906},
  url = {https://link.aps.org/doi/10.1103/PhysRevC.92.034906}
}

@article{Strangeness_enha_Nature,
   title={Enhanced production of multi-strange hadrons in high-multiplicity proton–proton collisions},
   collaboration = "ALICE collaboration",
   eprint = "1606.07424",
   archivePrefix = "arXiv",
   volume={13},
   ISSN={1745-2481},
   url={http://dx.doi.org/10.1038/nphys4111},
   DOI={10.1038/nphys4111},
   number={6},
   journal={Nature Phys.},
   publisher={Springer Science and Business Media LLC},
   author={Adam, J. and Adamová, D. and Aggarwal, M. M. and Rinella, G. Aglieri and Agnello, M. and Agrawal, N. and Ahammed, Z. and Ahmad, S. and Ahn, S. U. and Aiola, S. and Akindinov, A. and Alam, S. N. and Albuquerque, D. S. D. and Aleksandrov, D. and Alessandro,},
   year={2017},
   month=apr, 
   pages={535–539} }

@article{MultiStrangeBaryon,
title = {Multi-strange baryon production in p-{P}b collisions at $\sqsnn = 5.02\tev$},
collaboration = "ALICE collaboration",
archivePrefix = "arXiv",
eprint        = "1512.07227",
journal = {Phys. Lett.},
volume = {B758},
pages = {389-401},
year = {2016},
issn = {0370-2693},
doi = {https://doi.org/10.1016/j.physletb.2016.05.027},
url = {https://www.sciencedirect.com/science/article/pii/S0370269316301745},
author={Adam, J. and Adamová, D. and Aggarwal, M. M. and Rinella, G. Aglieri and Agnello, M. and Agrawal, N. and Ahammed, Z. and Ahmad, S. and Ahn, S. U. and Aiola, S. and Akindinov, A. and Alam, S. N. and Albuquerque, D. S. D. and Aleksandrov, D. and Alessandro,}
}

@article{BayesianUnfolding,
      title={Improved iterative {B}ayesian unfolding}, 
      author={G. D'Agostini},
      year={2010},
      archivePrefix={arXiv},
      eprint={1010.0632},
      primaryClass={physics.data-an},
      url={https://arxiv.org/abs/1010.0632}, 
}

@article{SHM_Nature,
    author = "Andronic, Anton and Braun-Munzinger, Peter and Redlich, Krzysztof and Stachel, Johanna",
    title = "{Decoding the phase structure of QCD via particle production at high energy}",
    eprint = "1710.09425",
    archivePrefix = "arXiv",
    primaryClass = "nucl-th",
    doi = "10.1038/s41586-018-0491-6",
    journal = "Nature",
    volume = "561",
    number = "7723",
    pages = "321--330",
    year = "2018"
}

@article{Flow_AuAu_STAR,
      author="Ackermann, K. H. and others",
      title="{Elliptic flow in {A}u+{A}u collisions at $\sqsnn = 130\; \mathrm{GeV}$}",
      collaboration="STAR collaboration",
      archivePrefix  = "arXiv",
      eprint         = "nucl-ex/0009011",
      primaryClass   = "nucl-ex",
      journal="{Phys. Rev. Lett.}",
      volume="86",
      year="2001",
      pages="402",
      doi="10.1103/physrevlett.86.402",
}

@article{CGC_flow1,
   title={Long range two-particle rapidity correlations in collisions from high energy {QCD} evolution},
   volume={A836},
   archivePrefix = "arXiv",
   eprint = "0911.2720",
   ISSN={0375-9474},
   url={http://dx.doi.org/10.1016/j.nuclphysa.2009.12.044},
   DOI={10.1016/j.nuclphysa.2009.12.044},
   number={1–2},
   journal={Nucl. Phys.},
   publisher={Elsevier BV},
   author={Dusling, Kevin and Gelis, François and Lappi, Tuomas and Venugopalan, Raju},
   year={2010},
   month=may, pages={159–182} }

@article{ALICE_mean_pt_2013,
   title={Multiplicity dependence of the average transverse momentum in pp, p–{P}b, and {P}b–{P}b collisions at the {LHC}},
   collaboration = "ALICE collaboration",
   archivePrefix = "arXiv",
   eprint = "1307.1094",
   volume={B727},
   ISSN={0370-2693},
   url={http://dx.doi.org/10.1016/j.physletb.2013.10.054},
   DOI={10.1016/j.physletb.2013.10.054},
   number={4–5},
   journal={Phys. Lett.},
   publisher={Elsevier BV},
   author={Abelev, B. and Adam, J. and Adamová, D. and Adare, A.M. and Aggarwal, M.M. and Aglieri Rinella, G. and Agnello, M. and Agocs, A.G. and Agostinelli, A. and Ahammed, Z. and Ahmad Masoodi, A. and Ahmad, N. and Ahmed, I. and Ahn, S.A. and Ahn, S.U. and Aimo},
   year={2013},
   month=dec, pages={371–380} }

@article{ALICE_XeXe_mean_pt,
   title={Multiplicity dependence of charged-particle production in pp, p–{P}b, {X}e–{X}e and {P}b–{P}b collisions at the {LHC}},
   collaboration = "ALICE collaboration",
   archivePrefix = "arXiv",
   eprint = "2211.15326",
   volume={B845},
   ISSN={0370-2693},
   url={http://dx.doi.org/10.1016/j.physletb.2023.138110},
   DOI={10.1016/j.physletb.2023.138110},
   journal={Phys. Lett.},
   publisher={Elsevier BV},
   author={Acharya, S. and Adamová, D. and Adler, A. and Aglieri Rinella, G. and Agnello, M. and Agrawal, N. and Ahammed, Z. and Ahmad, S. and Ahn, S.U. and Ahuja, I. and Akindinov, A. and Al-Turany, M. and Aleksandrov, D. and Alessandro, B.},
   year={2023},
   month=oct, pages={138110} }

@article{rcBK,
  title = {Nonlinear {QCD} meets data: A global analysis of lepton-proton scattering with running coupling {B}alitsky-{K}ovchegov evolution},
  author = {Albacete, Javier L. and Armesto, N\'estor and Milhano, Jos\'e Guilherme and Salgado, Carlos A.},
  journal = {Phys. Rev.},
  archivePrefix = "arXiv",
  eprint = "0902.1112",
  volume = {D80},
  issue = {3},
  pages = {034031},
  numpages = {11},
  year = {2009},
  month = {Aug},
  publisher = {American Physical Society},
  doi = {10.1103/PhysRevD.80.034031},
  url = {https://link.aps.org/doi/10.1103/PhysRevD.80.034031}
}

@techreport{ALICE_charged_part_def,
      title         = "{The ALICE definition of primary particles}",
      collaboration = "ALICE collaboration",
      author = "Acharya, S. and others",
      institution = "CERN",
      month         = "6",
      year          = "2017",
      number  = "{ALICE-PUBLIC-2017-005}",
      url           = "https://cds.cern.ch/record/2270008",
}

@article{ALICE_identified_charged,
   title={Measurement of pion, kaon and proton production in proton–proton collisions at $\sqs = 7\tev$},
   archivePrefix = "arXiv",
   eprint = "1504.00024",
   volume={C75},
   collaboration = "ALICE collaboration",
   ISSN={1434-6052},
   url={http://dx.doi.org/10.1140/epjc/s10052-015-3422-9},
   DOI={10.1140/epjc/s10052-015-3422-9},
   number={5},
   journal={Eur. Phys. J.},
   publisher={Springer Science and Business Media LLC},
   author={Adam, J. and Adamová, D. and Aggarwal, M. M. and Rinella, G. Aglieri and Agnello, M. and Agrawal, N. and Ahammed},
   year={2015},
   month=may }

@article{JY_flow_original,
  title = {Anisotropy as a signature of transverse collective flow},
  author = {Ollitrault, Jean-Yves},
  journal = {Phys. Rev.},
  volume = {D46},
  issue = {1},
  pages = {229--245},
  numpages = {0},
  year = {1992},
  month = {Jul},
  publisher = {American Physical Society},
  doi = {10.1103/PhysRevD.46.229},
  url = {https://link.aps.org/doi/10.1103/PhysRevD.46.229}
}

@article{CGC_flow2,
  title = {Initial-state geometry and the role of hydrodynamics in proton-proton, proton-nucleus, and deuteron-nucleus collisions},
  author = {Bzdak, Adam and Schenke, Bj\"orn and Tribedy, Prithwish and Venugopalan, Raju},
  journal = {Phys. Rev.},
  volume = {C87},
  issue = {6},
  pages = {064906},
  numpages = {10},
  year = {2013},
  month = {Jun},
  publisher = {American Physical Society},
  doi = {10.1103/PhysRevC.87.064906},
  url = {https://link.aps.org/doi/10.1103/PhysRevC.87.064906}
}

@article{JETSCAPE,
    author = "Mankolli, A. and others",
    collaboration = "JETSCAPE",
    title = "{Longitudinal Dynamics of Large and Small Systems from a 3D Bayesian Calibration of RHIC Top-energy Collision Data}",
    eprint = "2601.17234",
    archivePrefix = "arXiv",
    primaryClass = "nucl-ex",
    month = "1",
    year = "2026"
}

@article{GBW_model,
    author = "Golec-Biernat, Krzysztof J. and Wusthoff, M.",
    title = "{Saturation effects in deep inelastic scattering at low $Q^{2}$ and its implications on diffraction}",
    eprint = "hep-ph/9807513",
    archivePrefix = "arXiv",
    reportNumber = "DTP-98-50",
    doi = "10.1103/PhysRevD.59.014017",
    journal = "Phys. Rev. D",
    volume = "59",
    pages = "014017",
    year = "1998"
}

@article{CMS:2010ifv,
    author = "Khachatryan, Vardan and others",
    collaboration = "CMS",
    title = "Observation of Long-Range Near-Side Angular Correlations in Proton-Proton Collisions at the {LHC}",
    eprint = "1009.4122",
    archivePrefix = "arXiv",
    primaryClass = "hep-ex",
    reportNumber = "CMS-QCD-10-002, CERN-PH-EP-2010-031",
    doi = "10.1007/JHEP09(2010)091",
    journal = "JHEP",
    volume = "09",
    pages = "091",
    year = "2010"
}

@article{ALICE_NATURE_STRANGENESS,
   title={Enhanced production of multi-strange hadrons in high-multiplicity proton–proton collisions},
   volume={13},
   collaboration={ALICE collaboration},
   ISSN={1745-2481},
   url={http://dx.doi.org/10.1038/nphys4111},
   DOI={10.1038/nphys4111},
   number={6},
   journal={Nature Physics},
   publisher={Springer Science and Business Media LLC},
   author={Adam, J. and Adamová, D. and Adare, A.M. and Aggarwal, M.M. and Aglieri Rinella, G. and Agnello, M. and Agocs, A.G. and Agostinelli, A. and Ahammed, Z. and Ahmad Masoodi, A. and Ahmad, N. and Ahmed, I. and Ahn, S.A. and Ahn, S.U. and Aimo},
   year={2017},
   month=Apr, pages={535–539} }

@article{ALICE_fractions,
    author = "Adam, Jaroslav and others",
    collaboration = "ALICE",
    title = "{Multiplicity dependence of charged pion, kaon, and (anti)proton production at large transverse momentum in p-Pb collisions at $\sqsnn = 5.02 \tev$}",
    eprint = "1601.03658",
    archivePrefix = "arXiv",
    primaryClass = "nucl-ex",
    reportNumber = "CERN-EP-2016-003",
    doi = "10.1016/j.physletb.2016.07.050",
    journal = "Phys. Lett. B",
    volume = "760",
    pages = "720--735",
    year = "2016"
}

\newpage
% LHCb collaboration author list
% Data extracted on August 24th, 2026 at 11:15am for paper reference LHCb-PAPER-2026-012
\centerline
{\large\bf LHCb collaboration}
\begin
{flushleft}
\small
R.~Aaij$^{38}$\lhcborcid{0000-0003-0533-1952},
M.~Abdelfatah$^{69}$,
A.S.W.~Abdelmotteleb$^{57}$\lhcborcid{0000-0001-7905-0542},
C.~Abellan~Beteta$^{51}$\lhcborcid{0009-0009-0869-6798},
F.~Abudin\'en$^{59}$\lhcborcid{0000-0002-6737-3528},
T.~Ackernley$^{61}$\lhcborcid{0000-0002-5951-3498},
A.A.~Adefisoye$^{69}$\lhcborcid{0000-0003-2448-1550},
B.~Adeva$^{47}$\lhcborcid{0000-0001-9756-3712},
M.~Adinolfi$^{55}$\lhcborcid{0000-0002-1326-1264},
P.~Adlarson$^{87,42}$\lhcborcid{0000-0001-6280-3851}
% ...

{\footnotesize \it

% ...
$^{38}$Nikhef National Institute for Subatomic Physics, Amsterdam, Netherlands\\
% ...
$^{42}$National Center for Nuclear Research (NCBJ), Warsaw, Poland\\
% ...
$^{47}$Instituto Galego de F{\'\i}sica de Altas Enerx{\'\i}as (IGFAE), Universidade de Santiago de Compostela, Santiago de Compostela, Spain\\
% ...
$^{51}$Physik-Institut, Universit{\"a}t Z{\"u}rich, Z{\"u}rich, Switzerland\\
% ...
$^{55}$H.H. Wills Physics Laboratory, University of Bristol, Bristol, United Kingdom\\
% ...
$^{57}$Department of Physics, University of Warwick, Coventry, United Kingdom\\
% ...
$^{59}$School of Physics and Astronomy, University of Edinburgh, Edinburgh, United Kingdom\\
$^{60}$School of Physics and Astronomy, University of Glasgow, Glasgow, United Kingdom\\
$^{61}$Oliver Lodge Laboratory, University of Liverpool, Liverpool, United Kingdom\\
% ...
$^{69}$Syracuse University, Syracuse, NY, United States\\
% ...
$^{87}$Department of Physics and Astronomy, Uppsala University, Uppsala, Sweden, associated to $^{60}$\\
% ...
}
\end{flushleft} 

\end{document}